\documentclass[twocolumn,showpacs,preprintnumbers,amsmath,amssymb,letterhead]{revtex4}
\usepackage{graphicx}% Include figure files
\usepackage{url}
\newlength{\defbaselineskip}
\newcommand{\setlinespacing}[1]%
           {\setlength{\baselineskip}{#1 \defbaselineskip}}

\begin{document}

%to change 1) clarify slow pulling "undulation" statement
%2) explicitly point out over smoothed curve in fig 1
%3) include ray's concerns

\title{A Random Force is a Force, of Course, of Coarse:  \\ Decomposing Complex Enzyme Kinetics with Surrogate Models}

\begin{abstract}
The temporal autocorrelation (AC) function associated with monitoring order
parameters characterizing conformational fluctuations of an enzyme
is analyzed using a collection of surrogate models.
 The
surrogates considered are phenomenological stochastic differential
equation (SDE) models. It is demonstrated how an ensemble of such
surrogate models, each surrogate being calibrated from a single
trajectory, indirectly contains information about unresolved
conformational degrees of freedom. This ensemble can be used to
construct complex temporal ACs associated with a ``non-Markovian"
process. The ensemble of surrogates approach allows researchers to
consider models more flexible than a mixture of exponentials to
describe relaxation times and at the same time gain physical
information about the system.  The relevance of this type of
analysis to matching  single-molecule experiments to computer
simulations and how more complex stochastic processes can emerge
from a mixture of simpler processes is also discussed.  The ideas
are illustrated on a toy SDE model and on  molecular dynamics
simulations of the enzyme dihydrofolate reductase. 

  \end{abstract}

\author{Christopher P. Calderon $^\dagger$}

\email{calderon@rice.edu}
\affiliation{%
$^\dagger$ Department of Computational and
Applied Mathematics, Rice University, Houston, TX 77005, USA.
}

\date{ \today}

%PACS: 02.50.-r  87.10.Mn

\maketitle

%Potential Titles:
%-A Multiscale Approach to Modeling Complex Enzyme Kinetics in the Presence of Low Energetic Barriers.
%- Complex Enzyme Conformational Kinetics Modeled via a Randomized Mixture of Surrogate Models.
%-A  Random Force is a Force, of Course, of Coarse:   Decomposing Complex Enzyme Kinetics with Surrogate Models

%%%%%%%%%%%%%%%%%%%%%%%%%%%%%%%%%%%%%%%%%%%%%%%%%%
\section{Introduction}
When enzymes and other proteins are probed at the single-molecule level, it has been
 observed in both experiments \cite{xie_dynamicdis_98,xie04,Dobson_science06AdK, natureFRET07} and simulation
  studies \cite{wolynes_crackingPNAS03, AdK, wolynesHFSP08, arora09, rief_09,Portman_CaMcracking09} that conformational
  fluctuations at several disparate timescales have physically significant
   influence on both large scale structure and biochemical function.   In this article,
   a method for using a collection of Marovian surrogate
   models \cite{SPA1, SPA2,SPAJCTC}  to predict kinetics that would often be considered
non-Markovian is presented \cite{klafter08,kou_08}.  The ideas in
\cite{SPAJCTC}  are extended to treat a system with more complex
kinetics.   The aim of the approach is to obtain a better
quantitative understanding the factors contributing to  complex
time autocorrelations (ACs) associated with quantities modulated by
slowly evolving conformational degrees of freedom.
The focus is on systems  where certain thermodynamically important
conformational degrees of freedom evolve over an effective free
energy surface with relatively low-barriers; this situation is
often relevant to  molecules undergoing a  ``population shift" or
``selected-fit" mechanism \cite{AdK,arora09} and the connection to  ``dynamic disorder" \cite{xie_dynamicdis_98} is also discussed.  The particular enzyme studied is dihydrofolate reductase (DHFR) because of its biological relevance to therapeutics and also due to the rich kinetics observed in certain order parameters \cite{arora09}.

Surrogate models are used to describe the short time dynamics.  These surrogates are fairly
simple phenomenological parametric stochastic differential
equation (SDE) models.  Specifically  the  Ornstein-Ulhlenbeck
process and an overdamped Langevin
equation with a position dependent diffusion function
\cite{hummer_posdep, SPA1, wangPNAS07,SPAJCTC} are considered as the
candidate surrogate models. Position dependent diffusion is often
observed when a few observables (or order parameters) are used to describe an underlying
 complex system such as a protein.  Position dependent noise  models
 allow one to consider ACs having a different functional
 form than an exponential decay and it is demonstrated that this added
 flexibility can be of assistance in both understanding short and
 long timescale kinetics. % We would like to point out that  maximum likelihood type estimates \cite{llglassy} corresponding to transition densities, exact and approximate \cite{aitECO,ozaki},
 Maximum likelihood type estimates utilizing transition densities,
 exact and approximate \cite{aitECO,ozaki,chen09}, are used to fit our surrogate SDE models.
 The fitting method does not require one to discretize \cite{hummer_posdep} state space (the surrogates assume  a continuum of states).
  The temporal
AC is not used directly as a fitting criterion
\cite{kou_08,stuart_09}, but the surrogate models are able to
accurately predict the AC after the model parameters
are fit. Maximum likelihood based approaches employing accurate
transition density approximations and a parametric structure posses
several advantages in this type of application  \cite{llglassy}.
An accurate transition density of a parametric SDE facilitates
goodness-of-fit tests appropriate for both stationary
\cite{aitFAN05,chen08} and nonstationary time series \cite{hong}.
The latter is particularly relevant to  many systems (like the one
considered here) where the diffusion coefficient is modulated by
factors not directly monitored \cite{klafter08,SPAJCTC} and the
hypothesis of a fixed/stationary surrogate model describing the
modeled time series is questionable.
 Statistically testing the validity of various
assumptions explicitly or implicitly behind a candidate surrogate
model, such as Markovian dynamics, state dependent noise and/or
regime switching  is helpful in both experimental and simulation
data settings \cite{SPAfilter,SPAdsDNA}.

The type of modeling approach presented is attractive from a
physical standpoint  for a variety of reasons. The items that
follow are discussed further in the Results and Discussion:

\begin{itemize}
\item %(1)
In situations where the magnitude of the local fluctuations
depend significantly on the instantaneous value of the order
parameter monitored, simple exponential (or a finite mixture of
exponentials \cite{socci96}) can be inadequate to describe the
relaxation and/or AC function \cite{SPAJCTC}.  The surrogates
proposed can account for this situation when overdamped diffusion
models can be used; additionally the estimated model parameters
can  be physically interpreted.
\item %(2)
 It has been
observed that even in single-molecule trajectories that
``dynamic disorder" can be observed due to ignoring certain
conformational degrees of freedom
\cite{xie_dynamicdis_98,xie04,klafter08};  the methods  proposed here can be used to
account for this type of variability and  show promise
for comparing frequently sampled single-molecule experimental time
series to computer simulations where dynamic disorder is relevant.
\item % (3)
Changes in conformational fluctuation magnitudes
have been suggested to lead to physically interesting phenomena, so possessing a means for quantitatively describing an ensemble of dynamical responses can help  one in better understanding the complex dynamics of enzymes, e.g. \cite{wolynes_crackingPNAS03,AdK,wolynesHFSP08,Portman_CaMcracking09}.
 \item %  (4)
  There is
general interest in showing how more complex stochastic processes
arise from a collection of simpler parts \cite{granger80a,granger80b,cox91,klafter09}.
We discuss how, within a single trajectory,  a continuous type
of regime switching of Markovian surrogate models gives rise to an
AC that would be considered non-Markovian.
\end{itemize}

The remainder of this article is organized as follows:  Section
\ref{sec:bacmeth} reviews the background and presents the models  considered.  Section  \ref{sec:acmash}
  introduces
the  new modeling procedure used to approximate the AC function of
a molecule experiencing multiple types of fluctuations. Section
\ref{sec:results} presents the Results and Discussion, Section
\ref{sec:conc} provides the Summary and Conclusions and this is
followed by an Appendix.

%II Background and Methods
\section{Background and Methods}
\label{sec:bacmeth}
%intro general SDE and notation (JCTC copy)

\subsection{Effective  Dynamics and Statistical Inference}
%first talk about preparation of local time series.

The trajectory generated by a detailed molecular dynamics (MD) simulation will be
denoted by $\{ z_i \}_{i=1}^N$.  The dynamics of the order
parameter monitored  is assumed to be complex (nonlinear,
modulated by unobserved factors, etc.) even at the relatively
short $O(ns)$ time intervals the order parameter time series is observed
over.    However,  over short $\approx 50-100 ps$ time intervals  a continuous
stochastic differential equation (SDE)   having  the form
\begin{eqnarray}
 dz_t= & \mu(z_t; \theta,\Gamma)dt+ \sqrt{2}\sigma(z_t; \theta,
 \Gamma)dB_t,
\label{eq:SDEgeneric}
\end{eqnarray}
can often  approximate the effective stochastic dynamics of the order parameter \cite{SPA1,SPAJCTC}. In the above $\mu(\cdot)$ and $\sigma^2(\cdot) $ are the nonlinear
deterministic  drift and diffusion functions (respectively) and
$B_t$ represents the standard Brownian motion \cite{raoDIFF}.  The finite dimensional parameter vector is denoted by $\theta$  and $\Gamma$ is used to represent unresolved  lurking degrees of freedom that slowly modulate the dynamics \cite{gAIoan}.
%\footnote{The  label ``metastable" is not used  to describe the  situation where unresolved coordinates are %modulating the dynamics because sometimes the conformational coordinate slowly evolves/diffuses over a broad %region of phase space \cite{aroraAdK07,AdK} and defining a metastable set is problematic in such a situation.  %Furthermore, if large kinetic metastable barriers modulate the dynamics the methods presented in this article %would need to be modified.}

The surrogate SDE models are formed by first dividing each trajectory into
$L$ temporal partitions.  Each estimated parameter vector  is denoted by
$\theta_\ell$ using the sequence  $\{ z_i\}_{i=T_{\ell-1}}^{T_{\ell}}$ and an assumed model  , where  $\ell$ is an index of a
  partition, $1=:T_0< \ldots < T_\ell < \ldots < T_L := N$, used to
  divide a  time series into $L$ disjoint local temporal windows.  Within each of these windows, the data and the assumed model structure is used to compute $\theta_\ell $ using maximum likelihood type methods (exact \cite{chen09} and approximate \cite{aitECO} depending on the model).  The parametric structures considered are presented in the next section. It is to be stressed that we estimate a collection of models, $\{\theta_\ell\}_{\ell=1}^L$ for each trajectory.  The differences in the estimated parameters are due in part to random slowly evolving forces modulate the dynamics and also in part to unavoidable estimation uncertainty associated with a finite time series.  It is demonstrated that a collection of surrogate model parameter vectors is needed to summarize conformational fluctuations inherent to many complex biomolecules.  This procedure is repeated for each observed MD trajectory / time series.

% Hypothesis tests assessing the validity of
%the various assumptions behind this type of dynamical proxy can be employed to put this approximation on sounder footing %\cite{hong,aitFAN05,SPAJCTC}, and as  will be shown later can also determine when position dependent noise %\cite{hummer_posdep,gAIoan,SPAJCTC} becomes statistically significant.
The  term ``local diffusion coefficient" $\equiv
\widetilde{D}(z;\Gamma):= \sigma^2(z;\theta, \Gamma)$ is
introduced in order to distinguish the coefficient in the Eqn.
\ref{eq:SDEgeneric} from the diffusion coefficient usually implied
in the physical sciences: we estimate the former from observed
data.
  The term
``diffusion coefficient"   used in statistical physics
\cite{zwanzig} is not necessarily the same as
$\widetilde{D}(z;\Gamma)$.
If $\Gamma$ does not modulate the
dynamics, the two definitions are effectively identical. However one theme of this paper is that some  traditional dynamical summaries of  statistical physics, such as diffusion coefficient and ensemble based AC, can be modified or made less coarse by using a collection of surrogate models.  Such a procedure may help in intpretting/understanding single-molecule time series.

\subsection{Candidate Surrogate SDE Models}

Two local parametric SDE models considered.  MATLAB scripts illustrating how to obtain parameter estimates of both models from discretely observed data are available online \url{www.caam.rice.edu/tech_reports/2008_abstracts.html#TR08-25}.
  The first is a linear, constant additive noise process:
  \begin{equation}
       dz_t= B(A-z_t)dt + \sqrt{2}CdB_t.
   \end{equation}
   The  above SDE has a rich history in both the physical sciences \cite{gardiner} where it is usually referred to as  the Ornstein-Uhlenbeck (OU) process and in econometrics where it is sometimes referred to as the Vasicek process \cite{chen09}. The parameter vector to estimate in this model is $\theta \equiv (A,B,C)$.  This model is appealing for a variety of reasons, one being that the exact transition density and maximum likelihood parameter vector for a discretely sampled process \footnote{We assume here that the initial condition is not random and hence does not contribute to the log likelihood function.} can be written in closed-form, i.e.  a numerical optimization is not needed to find the parameter vector because the parameter estimate can be written explicitly in terms of $\theta$ and the observed data \cite{chen09}.

The second is
 a nonlinear, position dependent overdamped (PDOD) Langevin type SDE \cite{schultenJCP04,SPA1,SPAJCTC}:
  \begin{eqnarray}
       \nonumber dz_t= & \beta \Big(C+D(z_t-\psi_0)\Big)^2 \Big(A+B(z_t-\psi_0)\Big)dt \\
        & + \sqrt{2}\Big(C+D(z_t-\psi_0)\Big)dB_t.
        \label{eq:PDOD}
   \end{eqnarray}
The variable $\beta\equiv 1/(k_BT)$ is the inverse of the product of Boltzmann's constant and the system temperature. $\psi_0$ represents a free parameter; in this article it coincides with the umbrella sampling point specified in the simulation. % \ though other choices can be entertained \cite{SPA,gAIoan,SPAfilter}.
 The parameter vector to estimate in this model is $\theta \equiv (A,B,C,D)$.   Each parameter is estimated using the observed data and the
 transition density expansions \cite{aitECO} associated with
  Eqn. \ref{eq:PDOD}  is used to construct a log likelihood cost function. A  Nelder-Mead search is then used to find the $\theta$ maximizing the associated cost function.
The effective force  in the above model is assumed to be linear in $z$, e.g. $F(z):= A+B(z-\psi_0)$ whereas the diffusion function $\equiv \widetilde{D}(z;\Gamma):= \Big(C+D(z-\psi_0)\Big)^2$ is quadratic in $z$. The overdamped appellation comes from multiplying the effective force  by the effective friction (as determined by the Einstein relation \cite{schultenJCP04}) corresponding to this diffusion function.

 In this article, all stochastic integrals used are It\^{o} integrals.  When a  complex high-dimensional system with multiple timescales is approximated with a low dimensional SDE possessing position dependent noise the choice of the It\^{o} or Stratonovich  integral influences the   interpretation of the drift function and the issue of which interpretation is ``physically correct" is a nontrivial problem \cite{razstuart04,dima}.  A related item is the so-called ``noise-induced drift"  \cite{risken,arnold00}. Such a term is sometimes explicitly added to the drift \cite{arnold00}, one thermodynamic motivation for
  this is discussed  further in Section \ref{sec:standardcomp}.
  
    An appealing feature of the data-driven modeling procedure presented here and elsewhere \cite{SPA1,molsim,SPAfilter,SPAJCTC,gAIoan} is that various SDE models, of an explicitly specified form, can be considered, estimated, and tested using observed trajectories.  Statistical hypothesis tests making use of the conditional distribution (not just moments) of the assumed surrogate model can then be used to test
    if the model assumptions are justified for the observed data.   Tools from mathematical statistics \cite{hong,aitFAN05} facilitate quantitatively and rigorously
    testing  if certain features are required to adequately describe the stochastic dynamics.
     Many  features, e.g.  position-dependent noise,  would be hard to statistically check using  AC based  heuristic methods.
    Such heuristic checks   are traditionally used in statistical physics, e.g. \cite{bussi_08,dima}.

     The data-driven models are used to  approximate the stochastic evolution  of black-box data  and the estimated parameters do have a loose physical phenomenological interpretation.  If one desires to compute unambiguous physical quantities from  the estimated coefficients using a particular definition from  statistical physics, the models can also be used to generate data for this purpose.  For example,  surrogate models can generate   nonequilibrium (surrogate) work \cite{SPA1,SPA2,gAIoan} and, under various assumptions, a well-defined thermodynamic potential of mean force (PMF)  can be derived from such data \cite{hummerPNAS01,schultenJCP04}.
     This contrasts the case where one starts with a high dimensional stochastic process (of known functional form) and then uses stochastic analysis to reduce the  dimensionality of the system by first appealing to asymptotic arguments \cite{razstuart04} and then possibly modifying the resulting equations
     to achieve a desired physical constraint \cite{arnold00}.  In both analytical and data-driven cases, the  goal is often to construct a single limiting low-dimensional evolution equation that can be used to predict statistical properties of the complex system valid over longer time-scales \cite{schultenJCP04,razstuart04,dima,hummer_posdep}.  It is not quantitatively  clear at what timescale such an approximation (if any such useful approximation exists at all) is valid over.
     Furthermore  it is usually difficult to determine if an equilibrium concept such as a PMF connects simply to trajectory-wise kinetics in small complex systems experiencing fluctuations.   Again, an appealing feature of the approach advocated here is that various statistical hypothesis tests  \cite{hong,aitFAN05,chen08} can be used to quantitatively assess the validity of   proposed (reduced) evolution equations to see if physically convenient models are consistent with the observed data.  For example, such tests can be used to determine the time one needs to wait before inertia can be neglected  \cite{SPAJCTC}.

     Regarding the validity of using a \emph{single} surrogate SDE to approximate ``long term" $> O(ns)$ trends, a main underlying theme of this paper and others  \cite{gAIoan,SPA2,SPAJCTC,SPAfilter} is that the presence of a lurking slowly evolving degree of freedom, $\Gamma$, can significantly  complicate using a single  equation  and that methods for quantitatively accounting for this sort of variation are underdeveloped.  Information in these types of models have proven useful in both theoretical chemistry computations \cite{gAIoan,SPA2} and in characterizing nanoscale experimental data \cite{SPAdsDNA,SPAfilter,SPAfric}.
      Throughout this article, it is shown how the \emph{collection} of surrogate models can be linked with the ideas of ``dynamic-disorder" \cite{xie_dynamicdis_98} to make quantitative statements about systems observed at the single-molecule level.

\section{A Method for Computing the AC Function of Complex Systems}
 \label{sec:acmash}

Before providing the  algorithmic details of the method, the basic idea and motivation for the approach is sketched in words.  It is assumed that a $\Gamma$ type coordinate slowly evolves (diffusively) over a relatively flat region of an effective free energy surface.  This evolution  modulates the
stochastic dynamics of the order parameter modeled, e.g. it changes the diffusion coefficient function \cite{SPAJCTC,klafter08}.  However, due to the almost continuous nature of the parameter change, a sudden or sharp change in the process dynamics is assumed difficult to detect in short segments of the time series (sudden regime changes or barrier crossings are not readily apparent in the data).
 Over longer time intervals, the changes become significant and the validity of a simple SDE model like the ones considered here to describe the global dynamics become suspect.  However if the evolution rules are updated as time progresses in the spirit of a dynamic disorder description \cite{xie_dynamicdis_98}, then there is hope for using a collection of these models to summarize the dynamics.    Even if the data is truly stationary, some fluctuations due to a $\Gamma$ type coordinate may take a long time to be ``forgotten" \cite{xie04,kou_08}.
  The idea proposed here is to essentially to use the estimated model for a time commensurate with time interval length used for estimation/hypothesis testing and then suddenly switch model parameters.    By doing this, one can take a collection of fairly simple stochastic models and construct another stochastic process possessing a more complex AC function.

   One advantage of such a procedure is that an ensemble of  elementary or  phenomenological  pieces  can be constructed to gain a better understanding
 of how variation induced by slowly evolving fluctuations affects some system statistics and this information may help in better quantitatively
 understanding some recently proposed enzyme  mechanisms
 \cite{wolynes_crackingPNAS03, AdK, wolynesHFSP08, arora09, rief_09,Portman_CaMcracking09}.
This method is in line  with the single-molecule philosophy that dynamical details should not be obscured by bulk averaging artifacts when possible.
   It is demonstrated how using a traditional AC summary of the data would obscure information of this sort on a toy example.
     Since the timescales at which simulations and single-molecule experiments span are rapidly converging, this new type of dynamical summary can also be used to help in matching the kinetics of simulations and experiments and/or can be used to understand how more complex dynamics emerge from simpler evolution rules \cite{granger80a,granger80b,cox91,klafter09}.

 Now for the algorithm details. Recall that for a single trajectory coming from a high dimensional system, the time series data is divided into partitions  and within each partition the parameters of both candidate models are estimated by methods discussed in the previous section.   This  results in a collection  $\{\theta_\ell\}_{\ell=1}^L $   for a each trajectory observed.    
    The Euler-Maruyama \cite{kp} scheme is  used here to simulate $N_{MC}$ trajectories for each  $\theta \in \{\theta_\ell\}_{\ell=1}^L $.   The surrogate SDEs are recorded every $\delta t$  and denote  simulated order parameter time series by  $\big\{ {\{ z^{s,(j)}_{i} \}^{T_{\ell}}_{i=T_{\ell-1}}} \big\}_{\ell=1}^{L}$ with $j=1,\ldots N_{MC}$.
     To construct a new time series using the $N_{MC}$ trajectories generated, set $x^s=\{\},\ell=1$ and for t=1 to $n\times L$ repeat the following:

    \begin{enumerate}
    \setcounter{enumi}{0}
         \item  Draw Uniform Integer u $\in [1,N_{MC}]$.
      \item  Set $x^s=\Big\{x^s \ , \ {\{ z^{s,(u)} \}^{T_{\ell}}_{i=T_{\ell-1}}}   \Big\}$
      \item  Update counter $\ell=\mathrm{mod}(t,L)+1$
    \end{enumerate}

    The procedure described results in a new time series $\{x^s_i\}_{i=1}^{N^\prime}$ where $N^\prime\equiv n\times N$. Note that the time ordering of the original data is maintained and the last step forces the series $\{x^s_i\}_{i=1}^{N^\prime}$ to be periodic so time lags  $>N\delta t$ cannot be resolved with this method.  If the integer $n>1$, the series  $\{x^s_i\}_{i=1}^{N^\prime}$ contains more temporal samples than the original series.  A larger sample size reduces the statistical uncertainty in an empirically determined AC.  The issue of reducing uncertainty is subtle
     and is discussed in detail using the toy model presented in the next section.
     If the time ordering is believed irrelevant, the first step can be modified to drawing 2 random integers. The other random integer can be   used to randomize the $\ell$ index.   \footnote{Using a randomized $\ell$ index turns out to be adequate for DHFR (the accuracy gain of respecting the time ordering provided only marginal improvements), though we present the version respecting time ordering in the algorithm to simplify the exposition.}.

     This procedure can then be  repeated for each trajectory coming from a high dimensional system.
     It is to be stressed that suddenly and relatively infrequently regime switches (``barrier hopping") cannot be described with this method. If
       the simulation or experiment is associated with a system possessing a  jagged/rough free energy surface with many small barriers and if
       a single trajectory can frequently sample the hops, then there is hope for using this method.  However note that the method
       is designed to treat relatively smooth regime changes (i.e. regime changes hard to identify by simple visual inspection).
      A discussion on how the surrogate models can be potentially used in more complex situations is briefly discussed later.

%%%%%%%%%%%%%%%%%%%%%%%%%%%%%%%%%%%%%%%%%%%%%%

%also discuss long range dependence and nonergodic sampling (single-molecule experiments and simulations). regime switching complicates stationarity assumptions
%%%%%%%%%%%%%%%%%%%%%%%%%%%%%%%%%%%%%%%%%%%%%%

% III Results and Discussion
%IIIa Illustrative Example

%IIIb DHFR results and discussion

\section{Results and Discussion}
\label{sec:results}

\subsection{Toy Model}
In order to demonstrate the AC method  on a simple example and illustrate some points in a controlled setting we use the following  SDE model:
\begin{eqnarray}
                    dy^{\mathrm{I}}_t= & \kappa^0(\alpha^0-y^\mathrm{I}_t)dt + \eta^0 dB^1_t \\
\nonumber dy^{\mathrm{II}}_t= & \kappa^0(\alpha_t-y^\mathrm{II}_t)dt + \eta^0 dB^1_t \\
\nonumber dy^\mathrm{III}_t= & \kappa_t(\alpha_t-y^\mathrm{III}_t)dt + \eta_t dB^1_t \\
\nonumber d\alpha_t= & 1/\tau_0(\alpha^0-\alpha_t)dt + \sigma^{\alpha} dB^2_t \\
\nonumber d\kappa_t= & 1/\tau_0(\kappa^0-\kappa_t)dt + \sigma^{\kappa} dB^3_t \\
\nonumber d\eta_t= & 1/\tau_0(\eta^0-\eta_t)dt + \sigma^{\eta} dB^4_t ,
\end{eqnarray}

\noindent where the constants $\alpha^0,\kappa^0,\eta^0$ are meant
to play the role of the surrogate parameters $(A,B,C)$ in the OU
model. In the above expressions, superscripts are used simply to
distinguish different constants or processes and do not represent
exponentiation. Superscripts on the $dB_t$ terms  are used to
distinguish separate independent standard Brownian motions.
  The Roman numeral superscripts distinguish three cases: I) the standard OU model; II) an OU type model where the mean level, $\alpha$ evolves stochastically; and III) an OU type model where all parameters evolve stochastically.  The parameter $\tau_0$ dictates the time scale at which the OU parameters stochastically evolve. The evolution studied here is made to be slow relative to that dictated by $\kappa^0$.  The (assumed unobserved) processes $\alpha_t$ , $\kappa_t$ ,$\eta_t$ are meant to mimic a dynamic disorder  \cite{xie04} type
situation.

\begin{figure} [ht]  %0
\center
\includegraphics[angle=0,width=.235\textwidth]{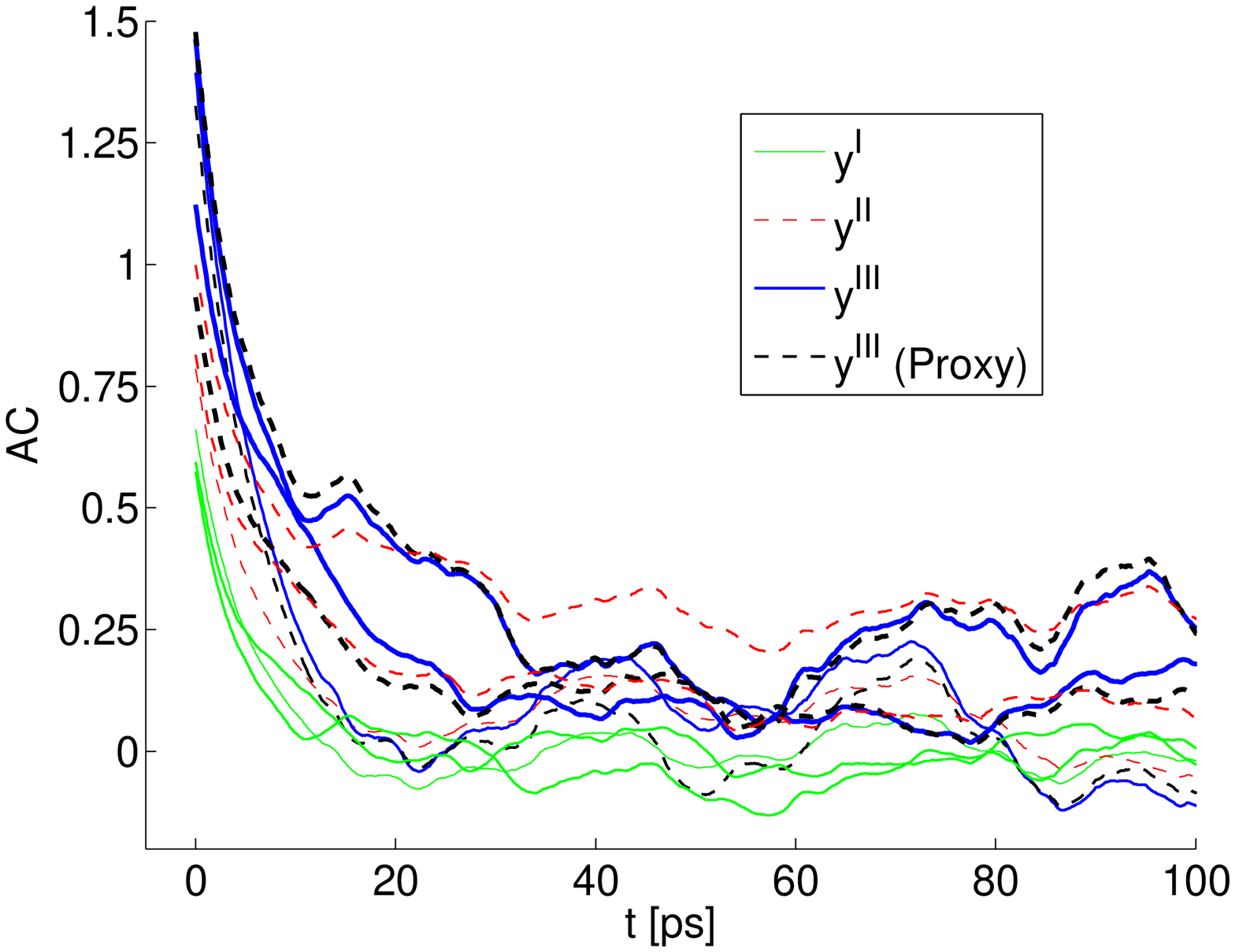}
\includegraphics[angle=0,width=.235\textwidth]{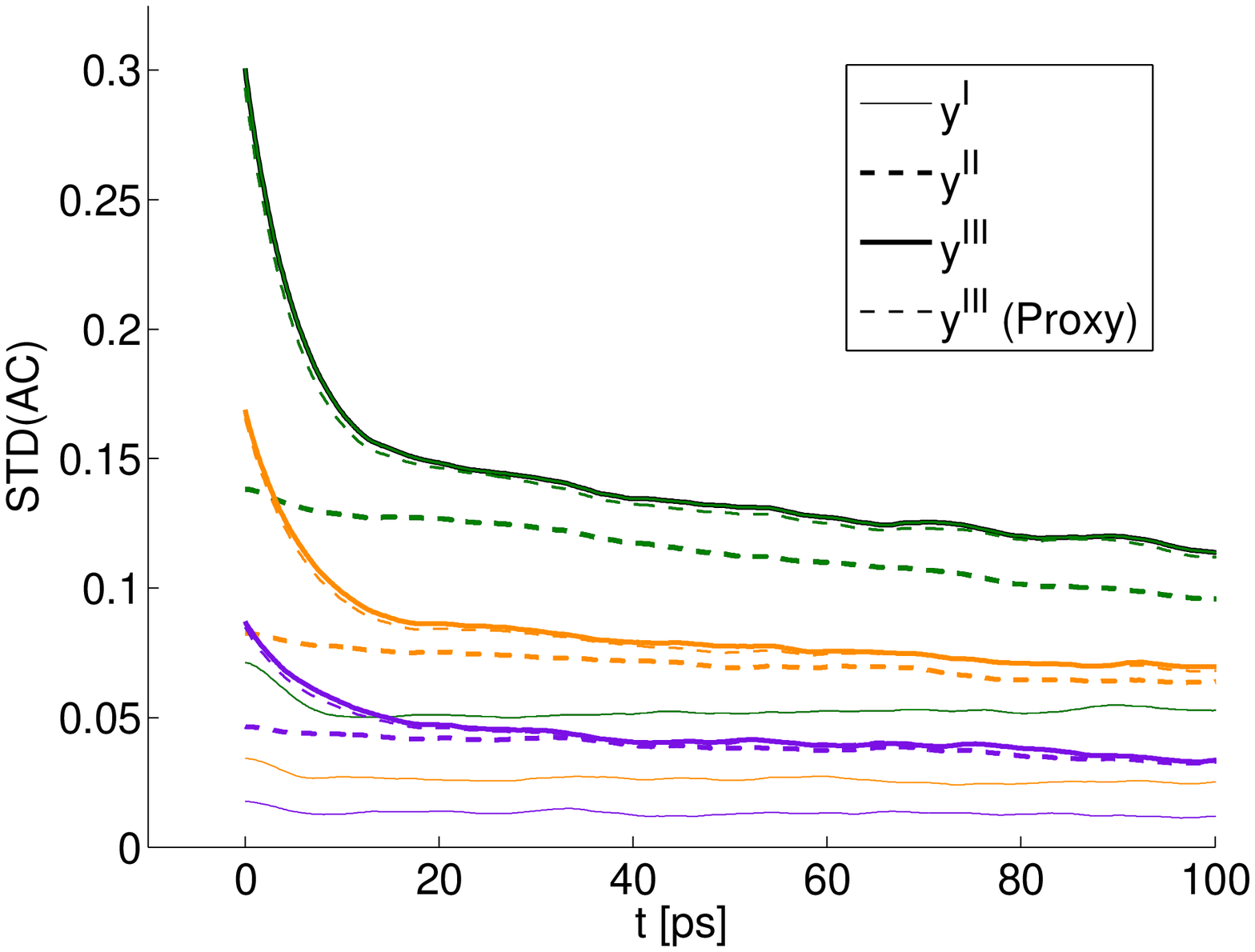}
\caption{   The AC of 4 realizations from 4 toy processes discussed in the text (left).  The standard deviation of the AC curves measured in a trajectorywise fashion
from a  population of 100 trajectories.  The computed standard deviation reflects  the  variance in a population of 100 empirically determined ACs.
  }
  \label{fig:1}
\end{figure}

  In addition, a fourth process  referred to as ``III (Proxy)" will be evolved to demonstrate the AC method of Section \ref{sec:acmash}.  This process is constructed by simply setting the parameters $\alpha^0,\kappa^0,\eta^0$ equal to the corresponding parameters of process III at time $t$ and then evolving this process like a standard OU model until the time index hits $t+\mathcal{T}$ when the parameters are updated to those of process III at the same time.  This procedure is then iterated. Randomizing $\mathcal{T}$ had little influence on the accuracy here, but can be entertained.
  % if the underlying effective free energy surface is believed to be very rugged.
    The processes above are simulated using the Euler-Maruyama scheme with time step size $\delta s$ and the process is observed discretely  every $\delta t$ time unit.
The remaining parameters are tuned to provide a parameter distribution consistent with those observed in some DHFR studies and  are reported in the Appendix.

The toy  model is used to investigate how  variation induced by slowly evolving $\Gamma$ type  factors influence the computed empirical AC on a controlled example where the assumptions behind the method introduced are satisfied.
  The features  discussed are relevant to the DHFR system studied later and are also likely relevant to oher single-molecules studies.     The example is also used to highlight issues relevant to nonergodic sampling \cite{klafter08}, i.e. when temporal averages are not equivalent to ensemble averages.  In this type of situation, single-molecule data is particularly helpful.
Use of the same Brownian motion term  to drive three separate processes facilitates studying contributions to variance in these types of studies.  In addition, estimation is not carried out to keep the discussion simple and to remove an additional source of uncertainty.

The left panel of Fig. \ref{fig:1} plots the empirical AC computed by sampling 4 realizations from this process using 5000 observations uniformly spaced by $\delta t=0.15 ps$.  These time series lengths are commensurate with those used in typical MD applications \cite{socci96,aroraAdK07,arora09}.  One observes that the slowly evolving parameters do influence the AC measured.  The fairly simple method of periodically updating the evolution parameters is able to mimic the AC associated with $y^\mathrm{III}$ for both short and long time scales.  Furthermore, the variation induced by the relaxation and noise level (modulated by $\kappa_t$ and $\eta_t$) influences both the short time and longer time responses.   The stochastic response of a dynamic disorder type process is  clearly richer than a single exponential. An advantage the surrogate approach offers over popular existing methods for treating this situation \cite{socci96} is that  other kinetic schemes, e.g. those associated with
overdamped models with position dependent diffusion, can be entertained.  In enzymes associated with complex dynamics,  other kinetics schemes may be needed to accurately reflect the stochastic dynamics of the order parameter monitored.  For example it is
demonstrated  in Fig. \ref{fig:3} that  the PDOD surrogate is needed accurately  captured relaxation kinetics even at  short $O(ps)$ timescales.  Over timescales relevant to experimentally accessible  order parameters    characterizing conformational fluctuations, one
 may need to account for  dynamical responses much richer than a mixture of exponentials  \cite{SPAJCTC,karplus08,kou_08}.    The procedure presented  demonstrated how
 ``elementary" pieces  could be patched together to characterize relaxations/fluctuations occurring over longer timescales.  This is attractive to both   computer simulations and experimental data sets.   In what follows the attention is shifted to focusing on limitations of using a single AC to describe single-molecule time series.

The right panel  of Fig. \ref{fig:1} plots the  standard deviation of the AC function associated with a  trajectory population.  For each observed trajectory, 100 different SDE trajectories were used to compute 100 empirical ACs from the time series associated with the trajectories.  The pointwise standard deviation measured over  the 100 ACs is plotted.
 The  curve shadings distinguish different time series sample sizes.  The three cases studied consisted of $(0.5,2,8)\times 10^4$ discrete temporal observations;
 each time series was uniformly sampled with $0.15 ps$ between observations.
  Note that  the influence of  the evolving parameters on the measured AC is substantial.  Recall all $y$ processes used common Brownian paths (so computer generated random numbers  do not contribute to the differences observed).  In addition, observe that  the difference between $y^{\mathrm{II}}$ and $y^{\mathrm{III}}$ persists for a fairly long time and the length of time this difference is measurably noticeable depends on the temporal sample size used to compute the AC.   In some applications, the variation induced by conformational fluctuations is important in computations \cite{gAIoan} or to characterize a system \cite{AdK,aroraAdK07}.
  The standard deviation in the measured AC here contains contributions coming from factors meant to mimic the influence
  unresolved conformational fluctuations whose influence persists for a fairly long time.  In the AC computed with longer time series, i.e. spanning a larger time since the time between observations is fixed, the process has more time to ``mix" and hence the difference between temporal and ensemble averages is reduced.  Said differently,  the influence of the initial conformation, or ``memory", diminishes.
  By using a single long time series trajectory and only reporting one AC computed from this ``mixed" series, these types of physically relevant fluctuations can get washed out by using a single AC function.    This goes against the spirit of single-molecule experiments.

The example considered here is admittedly simple and was constructed to illustrate the types of assumption behind the method introduced.  If
the dynamic disorder induced by large kinetic barriers or a complicated interaction with the surroundings, then one would need to construct  more sophisticated processes for determining how and when the parameters regime switch.
Combining the surrogate models with efforts along these lines, e.g. \cite{schuetteMMS09}, may be able to help these more exotic situations.
Exploring the various routes by which complex and/or heavy tailed ACs \cite{xie04,kou_08} can emerge from simpler dynamical rules
can help in a fundamental understanding of the governing physics \cite{granger80a,granger80b,cox91,klafter09}.  However, if the ensemble average
 decay rate is deemed the only  quantity of physical relevance then the collection of surrogate models can still potentially  be used to help in
 roughly predicting the rate of decay of more complex  ACs.  This is particularly relevant to simulations where
   obtaining long enough trajectories to reliably calibrate  models possessing complex AC exhibiting long range dependence
    from observed data is problematic \cite{Kaulakys_06, SPAJCTC,kou_08}.
 Even in  cases where one only requires  the asymptotic time decay of an ensemble of conformations for a physical  computation \cite{deshaw} and can simulate for a long enough time to directly monitor kinetics, an understanding of the distribution of surrogate models estimated will likely be of help in linking
   computer simulation force fields to  single-molecule experimental time series.     The remaining results use simulations of DHFR to illustrate some of these points.

\subsection{Dihydrofolate Reductase(DHFR)}

\begin{figure} [ht]  %0
\center
\includegraphics[angle=0,width=.45\textwidth]{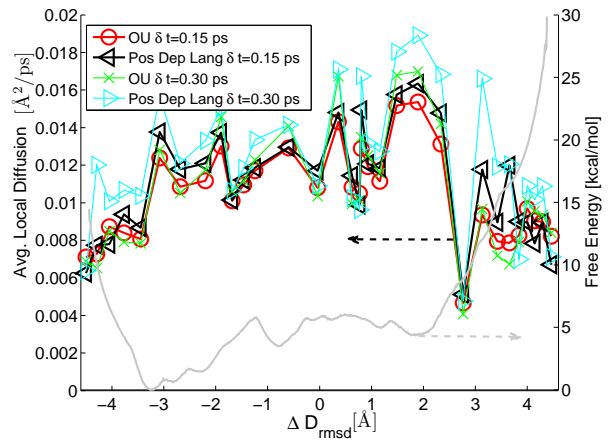}%
\caption{   Average local diffusion function estimated (left axis) and free energy (right axis) as a function of order parameter.  The population average of the estimated $C$ is plotted to give a feel for the position dependence of this quantity; it is stressed that the average alone is not adequate the describe the dynamics here due to a type of ``dynamic disorder" phenomena \cite{xie_dynamicdis_98}.  The free energy was computed by K. Arora using methods described in Ref. \cite{arora09}.
  }
  \label{fig:2}
\end{figure}

\subsubsection{DHFR Simulation Details}

The  detailed computational details are reported in Ref. \cite{arora09}.  Briefly,  an  order parameter denoted by $\Delta D_{\mathrm{rmsd}}$,
was defined using the root mean square distance between two crystal structures \cite{arora09}. This order parameter
 provides an indication of the proximity to the ``closed" and ``occluded" enzyme state and is reported in units of $\mathrm{\AA}$ throughout.
 The initial path between the closed and
occluded conformations of DHFR was generated using the Nudged Elastic Band (NEB) method~\cite{chu03}. Subsequently, $\approx $50
configurations obtained from NEB path optimization were subjected to US simulations.
During these US simulations, production dynamics of 1.2$ps$ at 300K
was performed after equilibration using a weak harmonic  restraint.

\subsubsection{DHFR Results}

Figure \ref{fig:2} plots the average local diffusion coefficient of the  surrogate SDE models using two different observation frequencies on the left axis and on the right axis the free energy computed in Ref. \cite{arora09} is plotted. Each surrogate model was estimated using 400 time series observations with either $\delta=0.15$ or $\delta=0.30 \ ps$  separating adjacent
 observations corresponding to $L=20$ or $L=10$ (respectively).    The average local diffusion coefficient demonstrates
a relatively smooth increasing trend for a majority of the order parameter values explored, but then suddenly changes abruptly around $\Delta D_{\mathrm{rmsd}} \approx 3 \mathrm{\AA}$.  It has been observed that an interesting interplay between free energy, fluctuations and stiffness, exists in some enzyme systems \cite{wolynes_crackingPNAS03,AdK,wolynesHFSP08,Portman_CaMcracking09} and this plot suggests that future
works  investigating some of the finer structural factors leading to these change may be worthwhile, though this direction is left to future work because it is outside the scope of this study.

\begin{figure} [ht]  %0
\center
\includegraphics[angle=0,width=.225\textwidth]{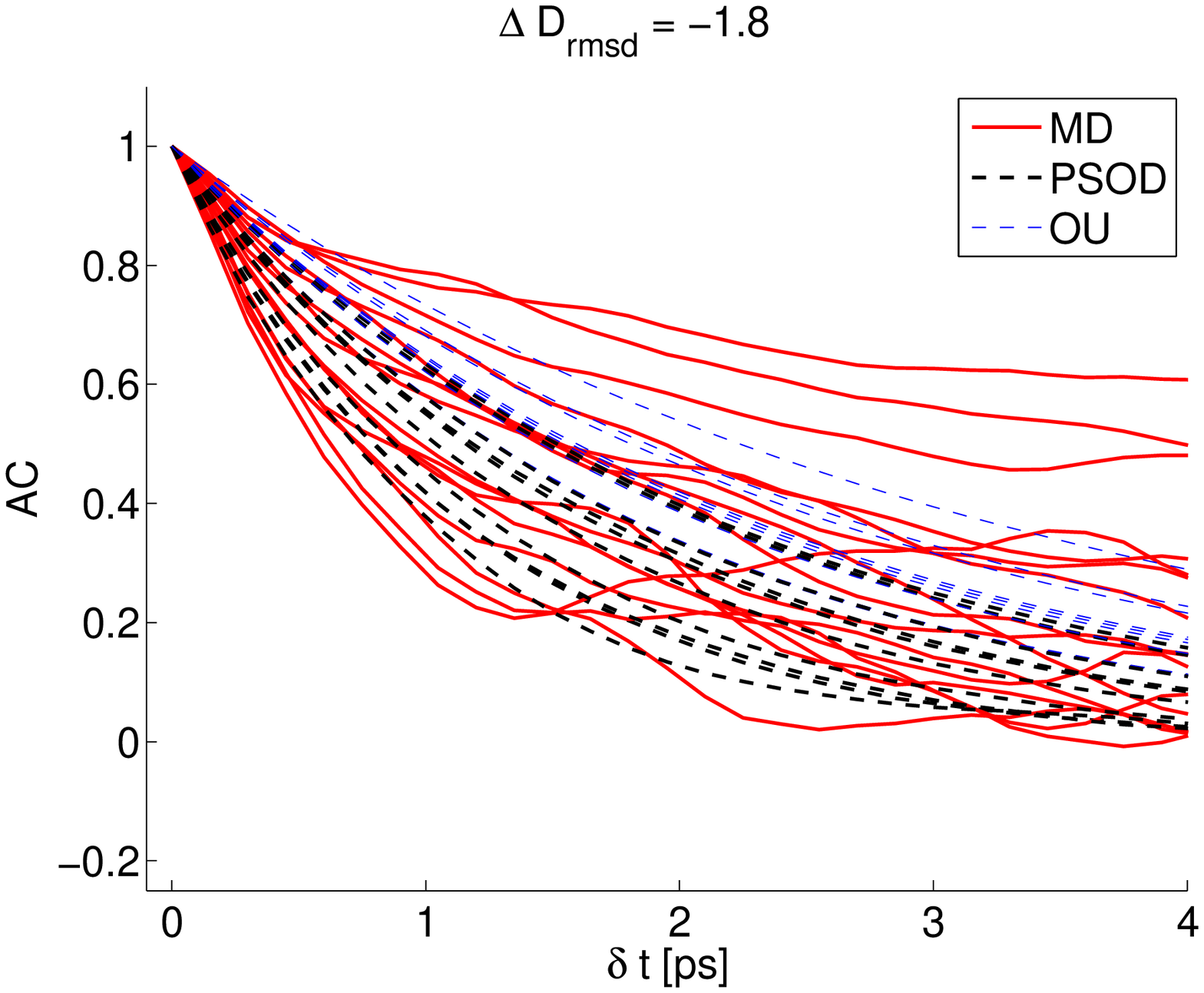} %
\includegraphics[angle=0,width=.225\textwidth]{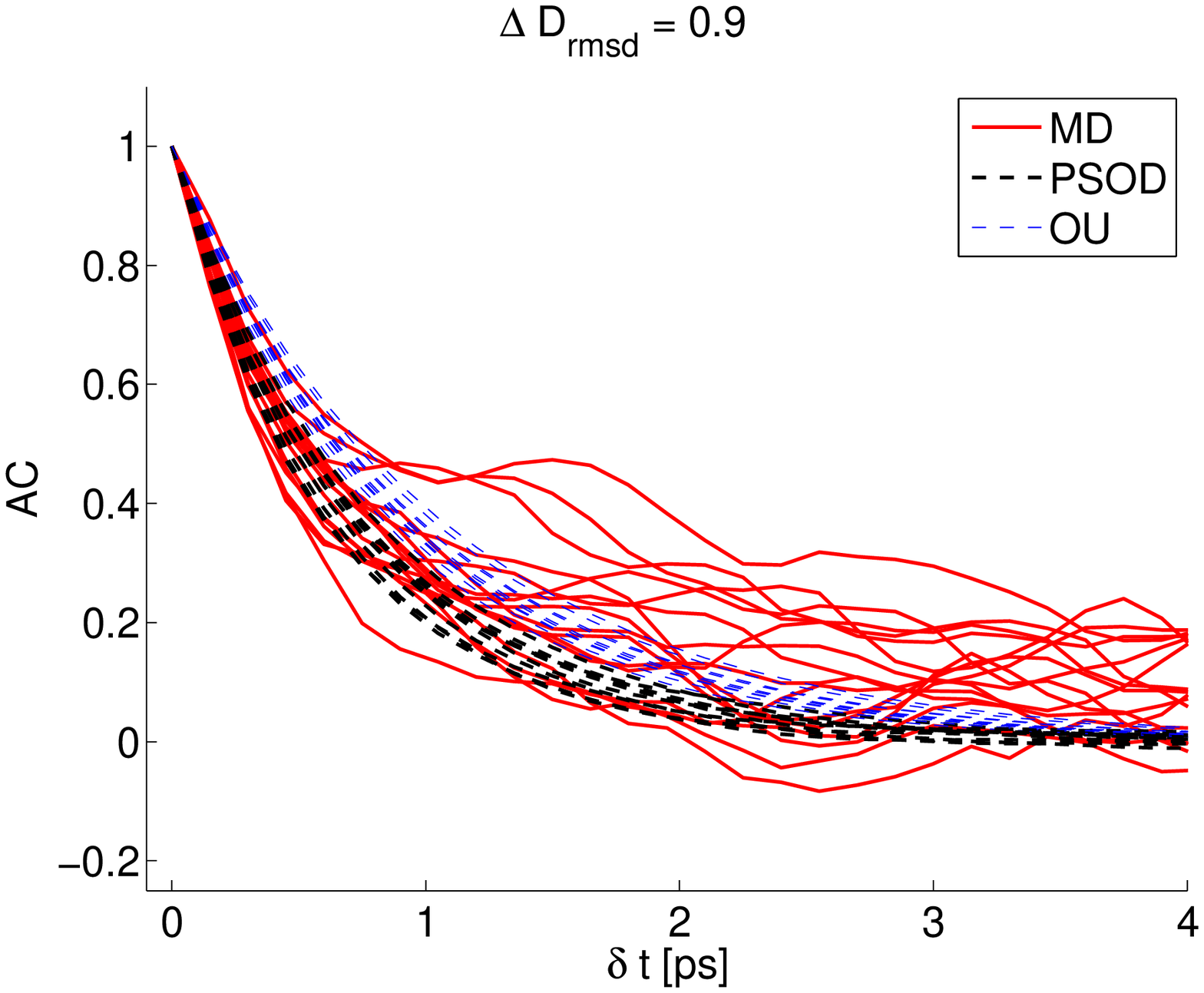} %
\includegraphics[angle=0,width=.225\textwidth]{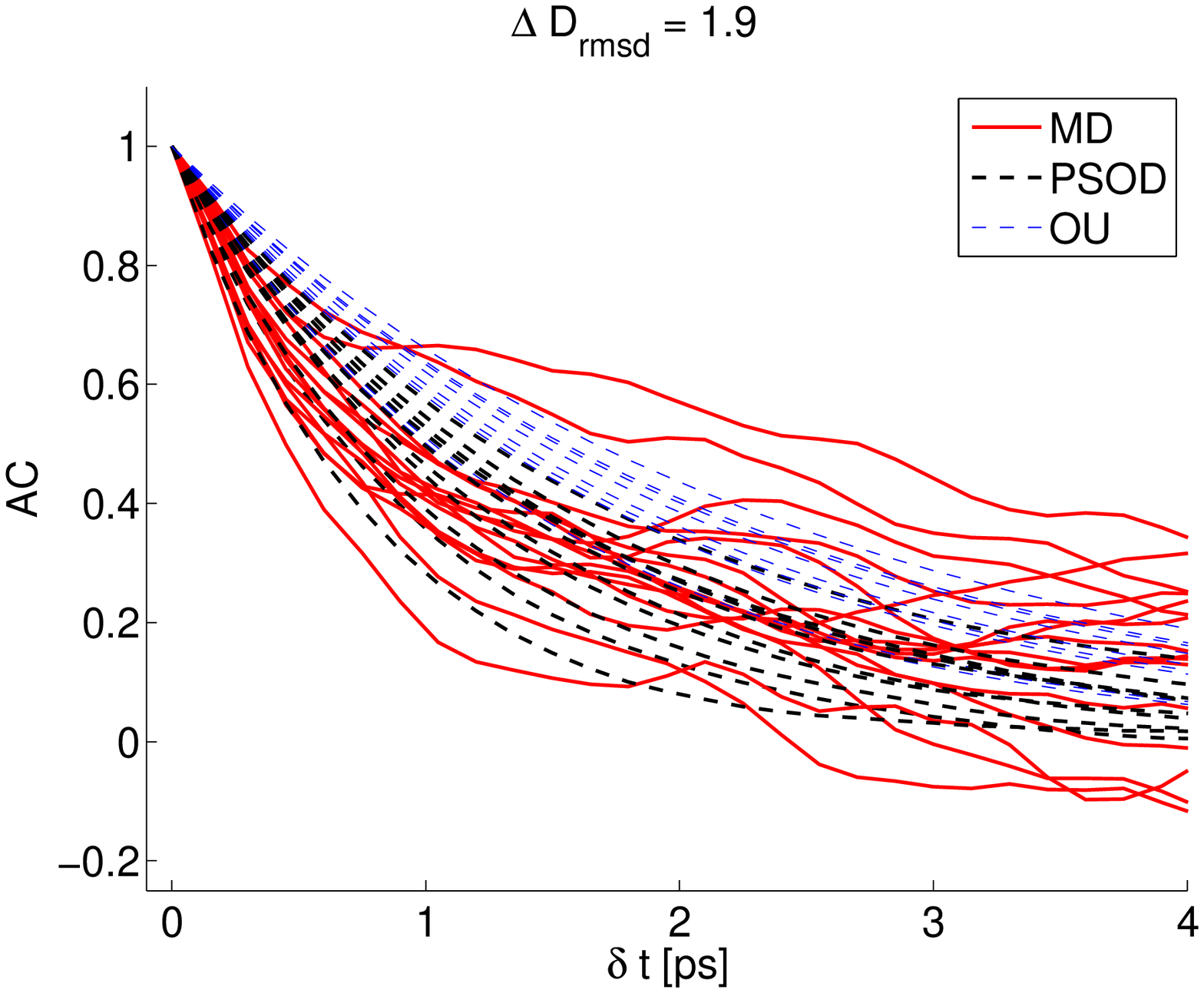} %
\includegraphics[angle=0,width=.225\textwidth]{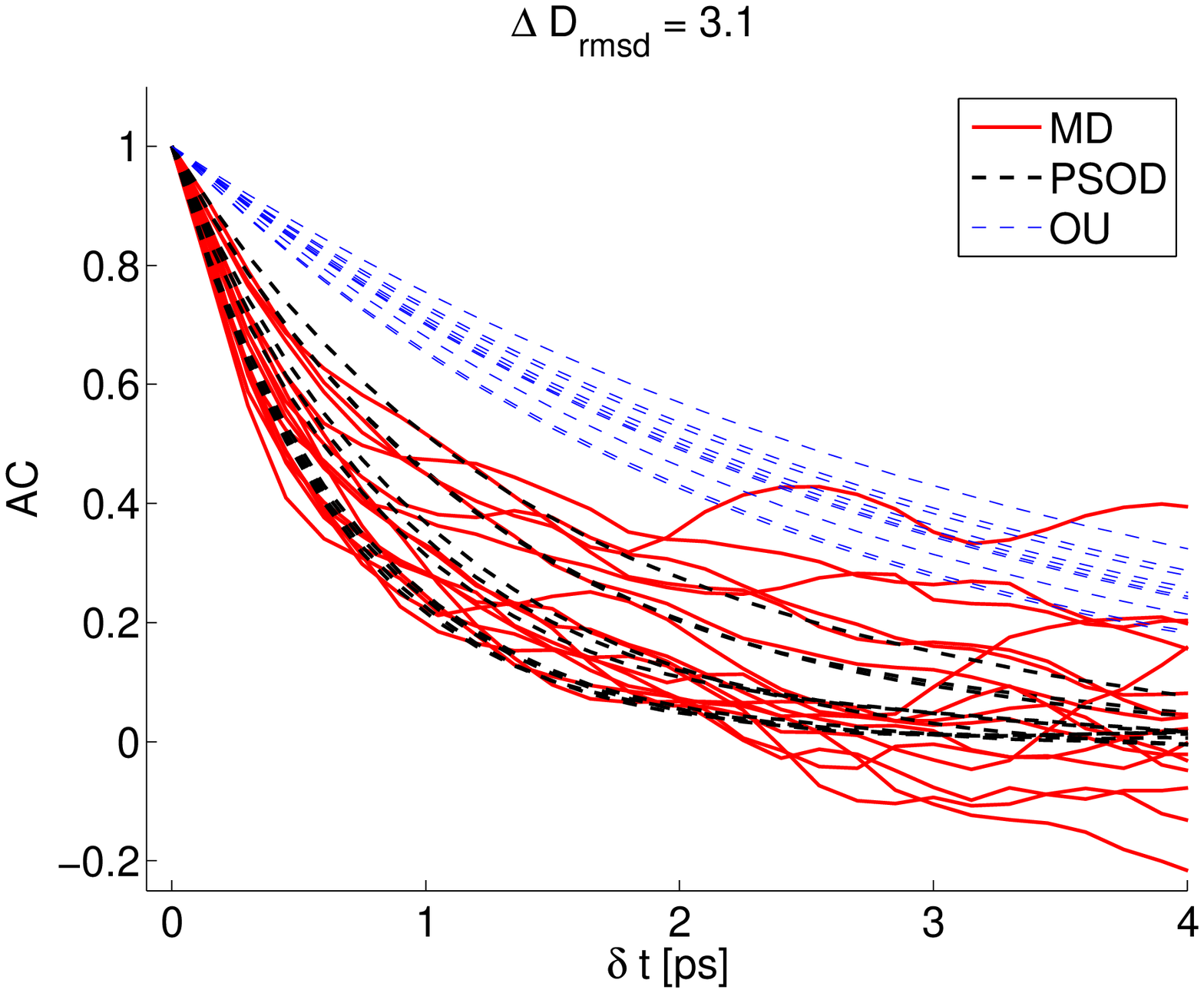} %
\caption{   Local AC Function of MD data and that predicted by surrogates.  The total time series of length 1.2 ns was divided into separate blocks, each containing 400 observations spaced by 0.15 ps.   This data was used to estimate a collection of surrogate models and a collection of MD ACs corresponding to US target points of $\Delta D_{\mathrm{rmsd}}\approx -1.8, 0.9, 1.9$ and $3.1 \mathrm{\AA}$.  The  AC corresponding to the estimate surrogate models is also plotted where the initial lag is normalized to unity to facilitate comparison (the raw units are shown in Fig. \ref{fig:3}).
  }
  \label{fig:3}
\end{figure}

\begin{figure} [ht]  %0
\center
\includegraphics[angle=0,width=.225\textwidth]{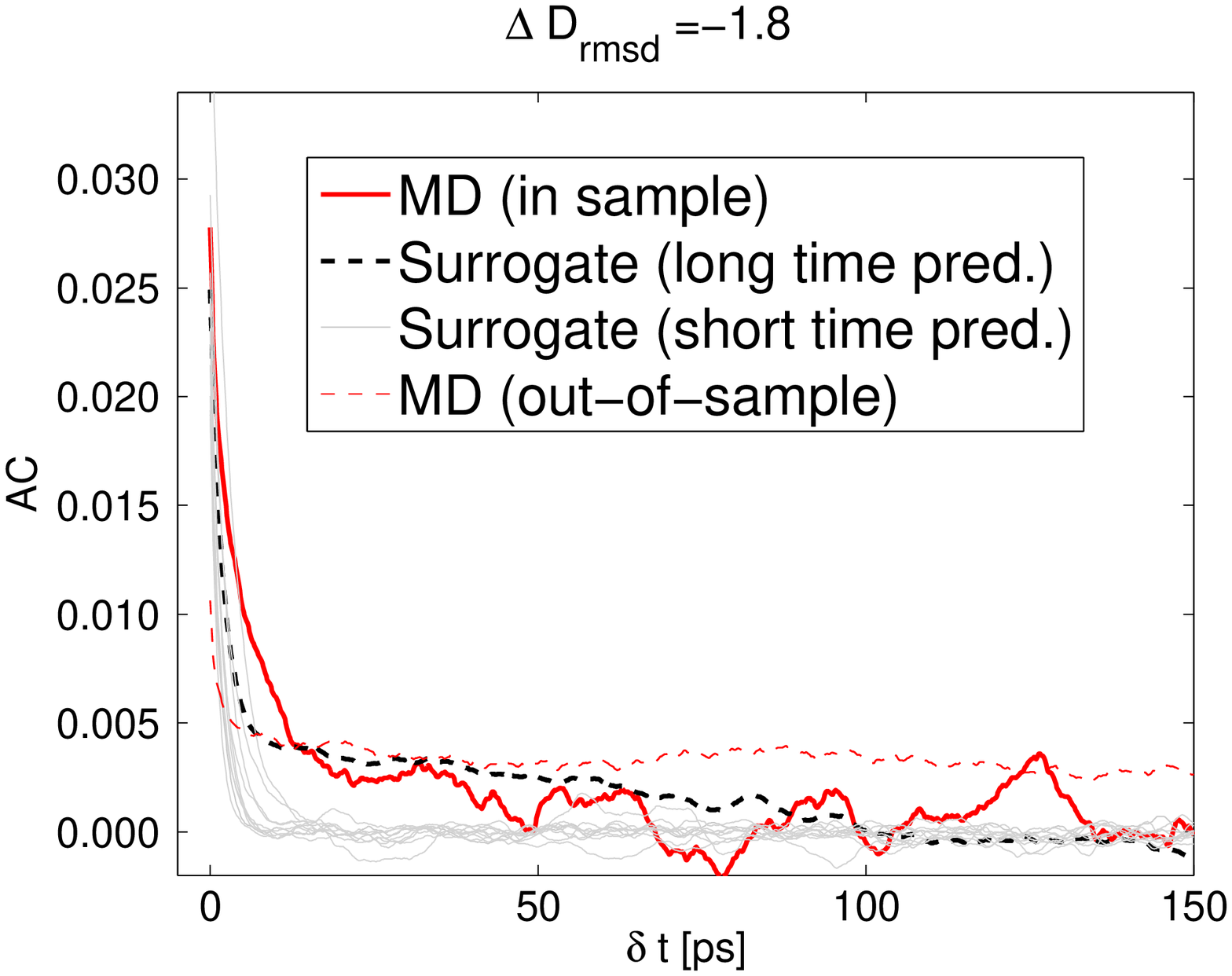} %
\includegraphics[angle=0,width=.225\textwidth]{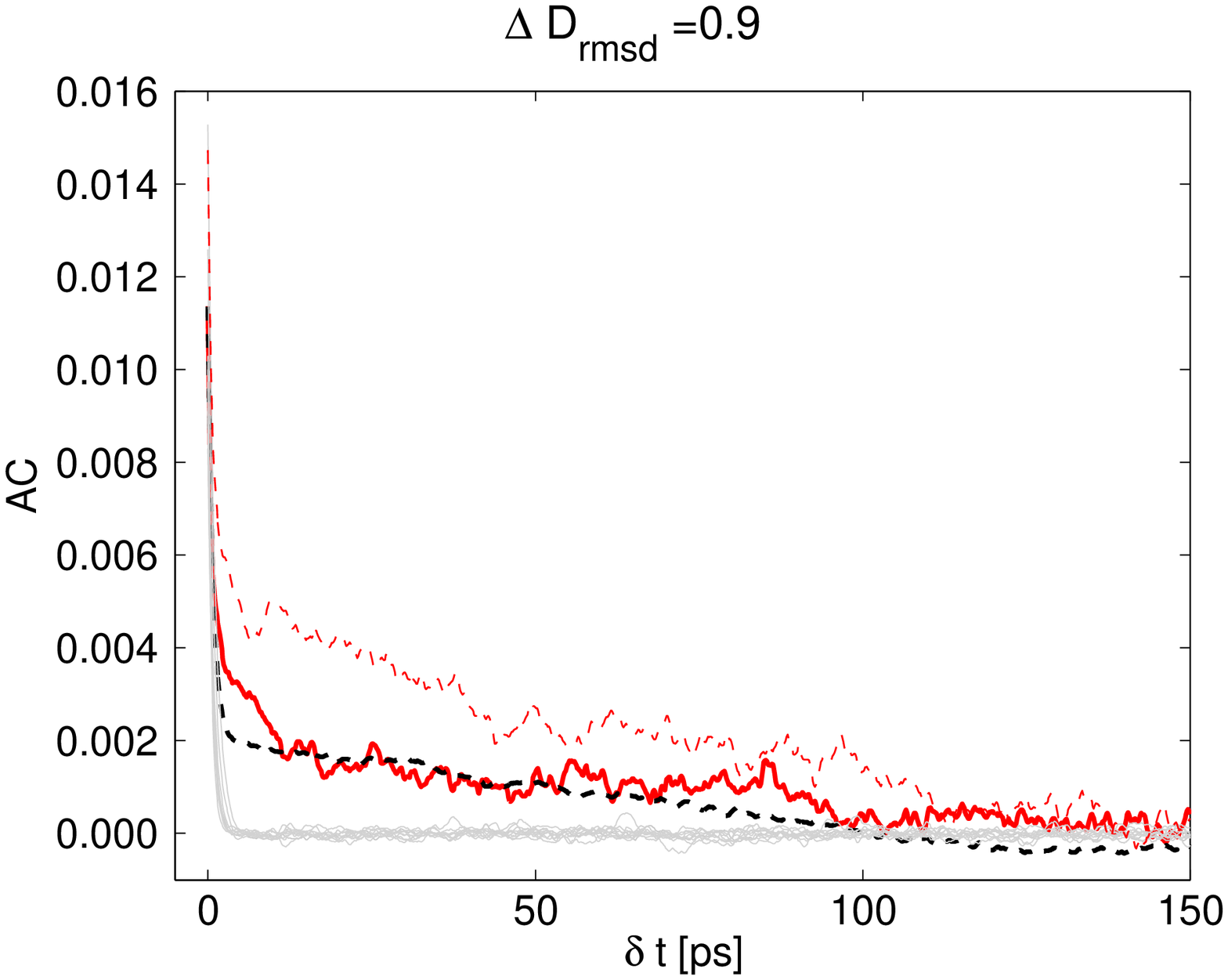} %
\includegraphics[angle=0,width=.225\textwidth]{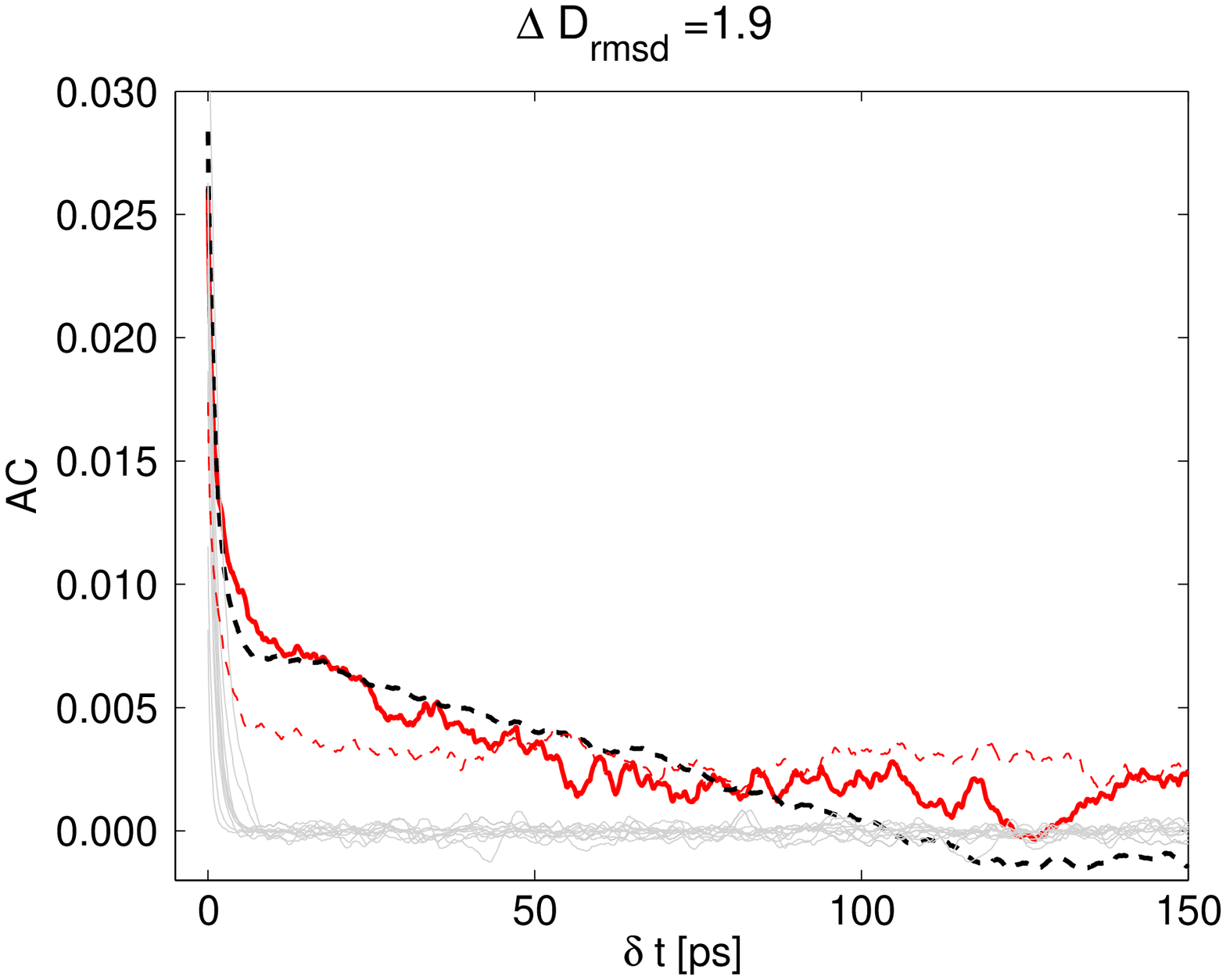} %
\includegraphics[angle=0,width=.225\textwidth]{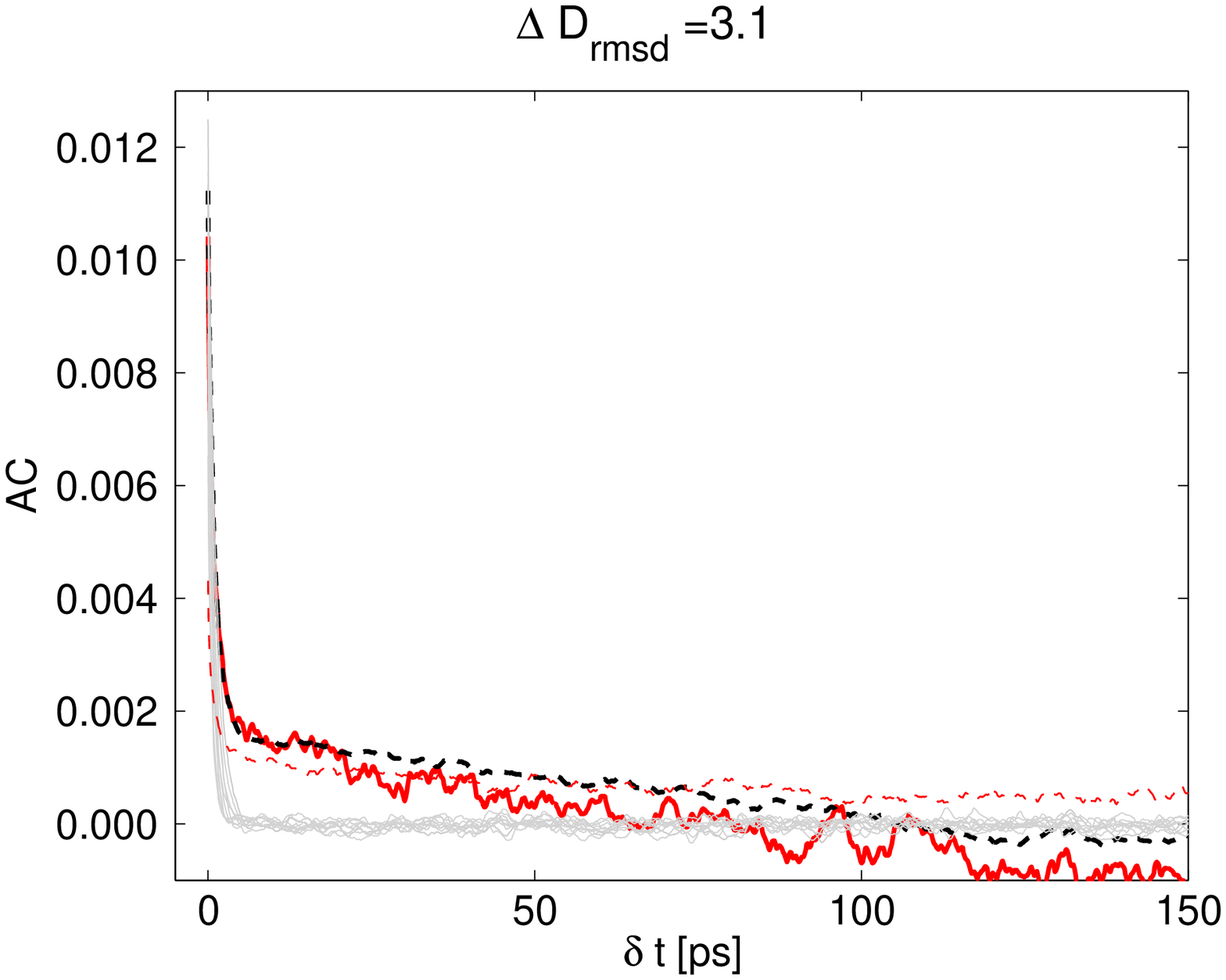} %
\caption{   The global AC prediction.  The procedure introduced here is used along with PDOD data to predict the long time AC.  The relaxation
predicted by the individual surrogate (data shown in Fig. \ref{fig:2} is also shown to stress that the SDE parameters are not fixed, but evolving and any single surrogate cannot capture the richer long time dynamics.
  }
  \label{fig:4}
\end{figure}

\begin{figure} [ht]  %0
\center
\includegraphics[angle=0,width=.45\textwidth]{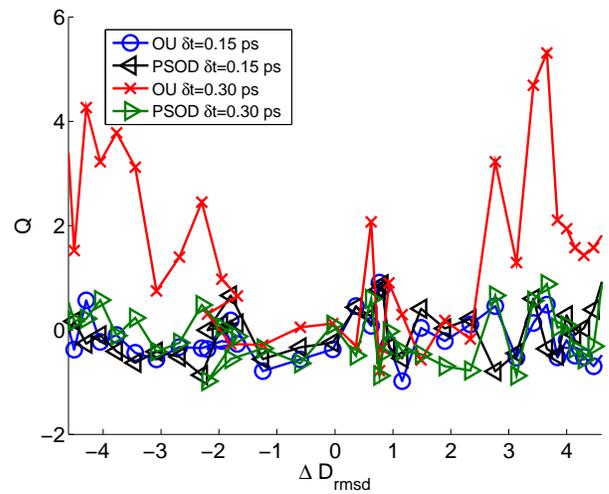}
\caption{   Goodness-of-fit tests. The test of Ref. \cite{hong} was computed given the parameter estimate and the observed data.  The median of each umbrella sampling window is reported.
  }
  \label{fig:5}
\end{figure}

It is to be stressed that the mean of each US window is not adequate to summarize the dynamics.  That is, a single fixed parameter surrogate SDE like the ones considered here cannot mimic the longer time statistics of the process. This is why the AC procedure introduced in Section \ref{sec:acmash} is needed.  Figure  \ref{fig:3} demonstrates that the individual PDOD models do capture features simpler surrogates cannot.  This is  due part to the position dependence of the local diffusion coefficient.  The PDOD surrogate model combined with the procedure of Section \ref{sec:acmash} can accurately summarize the long time dynamics.  These points are  explained further in the  discussion associated with Figs. \ref{fig:3}-\ref{fig:5}.

The ability of the PDOD model to capture features that a single
exponential (e.g. the AC associated with an OU process) cannot is
demonstrated in Fig. \ref{fig:3}.  Results from four different US
points, each possessing a different degrees of position dependence
on the noise are shown.  Here the results obtained using     both
the OU and PDOD surrogates calibrated using the  $\delta t=0.15
ps$ with 400 temporal observations  and the corresponding AC
predictions are shown in the plot.   The empirical ACs computed
using the short segments of MD data used for surrogate model
parameter estimation are also reported.  Results with 400 blocks
possessing observations spaced by $\delta t=0.30 ps$ were similar
in their AC prediction, but hypothesis tests strongly rejected the
assumption of a fixed local diffusion (see Fig \ref{fig:5}).  The
400 $\delta t=0.15 ps$ samples allowed the OU model structure to
provide a better  fit (as measured the fraction rejected)  because
the local diffusion function had less time to evolve/change value.
For cases where the position dependence is moderate,
 the PDOD and OU surrogate models
 predict qualitatively similar AC functions.
 However, the PDOD model captures the short time relaxation dynamics better than the OU for cases where the position
 dependence of the local diffusion is more substantial and hence for clarity we focus on the PDOD models in the remaining kinetic studies.

Figure \ref{fig:4} plots the empirically determined AC obtained
from different MD production simulation data.  The case labelled
``in-sample" was the one used for estimation of the local models
reported in Fig. \ref{fig:3} and that labeled ``out-of-sample" was
computed by running a longer 3.6 ns simulation and computing the
AC from the last 1.2 ns of this time series. The PDOD version of
these models were used along with the procedure outlined in
Section \ref{sec:acmash} using blocks of size 800 and randomizing
the time index.   The 400 blocks results were similar.  Respecting
the time ordering of the surrogate models only improved results
marginally.
  Note also that the general trends of the long time decay of the MD data is  captured with the procedure and that there is substantial difference between  the ``in-sample" and ``out-of-sample" MD trajectories \footnote{The large differences persist even if the time series length is increased by a factor of
  3.}.
    The physical relevance of such variation  was previously discussed and will be expanded on
   when results of  stationary DHFR density prediction are shown.
   The primary observation  is that a \emph{collection} of PDOD surrogate models were able to capture the basic relaxation trends
   of the enzyme that a single surrogate could not.  Recall that even at short timescales a single exponential decay was inadequate to
    fit the data. Similar trends were observed for all 51 US windows explored.   However, it is to be stressed that the
    procedure shown here is to \emph{decompose} kinetics  in the longest contiguous block of discrete time series observed.  If  complex
    dynamics occur over longer timescales and data is not available that directly samples these scales, then the method  cannot be used to predict
    the long time behavior that was not sampled.

\begin{figure} [ht]  %0
\center
\includegraphics[angle=0,width=.45\textwidth]{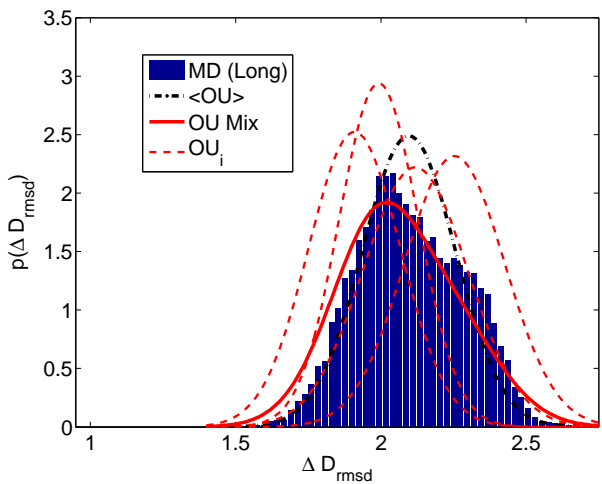}
\caption{   Stationary density/histogram prediction.  The bars denote 1.2 $ns$ MD data, the solid thick line to the mixture PDOD method (see text), the solid dotted line to the average PDOD model, and the thin lines to the 4 local equilibrium densities used in constructing the mixture density.
  }
  \label{fig:6}
\end{figure}

The goodness-of-fit of the surrogate models using the two candidate SDEs is shown in Fig. \ref{fig:5} for various US windows.  The median of the Q-test statistic introduced by \cite{hong} is reported.   This test statistic under the null is asymptotically  normally distributed with mean zero and unit variance, but has also been proven to be useful in small samples  \cite{hong,molsim, SPAJCTC,gAIoan}.  Recall that each MD time series (at each umbrella sampling window) was divided into small pieces.  In the portions near the edges (larger $|\Delta_{\mathrm{rmsd}}|$ values), where the position dependence of the noise is greatest, one observes that the OU model population has a median that would typically indicate  a collection of poor dynamical models.  If conformational fluctuations slowly modulate the dynamics, the  longer the time series one has, the likelihood of  departing from any simple surrogate model increases \footnote{As always, when the amount of evidence increases, the likelihood of rejecting any over simplified model increases.  However the tests developed in Ref. \cite{hong} are associated with ``diagnostics" which can help one in determining if a rejected model might still contain useful information nonetheless.}.  Goodness-of-fit tests, like the one presented here, can be used to quantitatively approximate when simple models begin  departing from various  assumptions.

The stationary density predicted by the surrogate OU models in a case where position-dependence was shown to be marginal for the time interval data was monitored is plotted in Fig. \ref{fig:6}.  Here the mixture method discussed in Ref. \cite{SPAJCTC} is reported due to its relevance to a collection of surrogates and dynamic disorder.  The histogram of the $1.2 ns$ MD data is also plotted as  well as the stationary density predicted by a the average of the surrogate models taken at the
 US window near $\Delta D_{\mathrm{rmsd}}\approx 2$.  Using a single model obtained by aggregating all time series together in hopes of reducing surrogate parameter uncertainty due to the resulting smaller time series sample sizes actually worsens the results.  The mixture of OU models calibrated using 4 sets of noncontiguously spaced  $60 ps$ data (i.e. $ 400$ entries spaced $0.15 ps$) were sampled every 300 ps from the MD process and this was used to compute 4 surrogate OU model parameters.
The goodness-of-fit tests indicated that the local surrogates
given the data were reasonable dynamical models.  So portions
where the ``local equilibrium" density, i.e. the stationary
density predicted by a surrogate with estimated parameters,
possessing significant probability mass can be though of the
regions of phase space sampled due to  fast-scale motion for a
relatively fixed (and unobserved) value of $\Gamma$
\cite{SPAJCTC,SPAfilter}. If variation in the conformational
coordinate is important to thermodynamic averages, as the data
here suggests to be the case in DHFR, then one needs to use a
collection of ``local equilibrium" densities \cite{SPAJCTC}. The
advantage of such an approach is that short bursts of simulations
started from different initial conditions can be run, then
surrogate models can be calibrated and tested. If the surrogate is
found suitable, it can then be used to make predictions on the
local equilibrium density, and the variation in the local
equilibrium densities  can be used to partially quantify the
degree to which a slow conformational degree of freedom modulates
the dynamics.
%The histogram result is shown here to illustrate a consequence of dynamic disorder on a thermodynamic quantity.
This treatment is appealing when data on other physically relevant order parameters is unknown are not easy to access.

%IV Conclusions
\section{Summary and Conclusions}
\label{sec:conc}

 Single-molecule experiments and simulations  offer the potential for  a detailed  fundamental understanding of complex biomolecules  without artifacts of bulk measurements obscuring the results.  However, one must deal with complex multiscale fluctuations at this level of resolution and the factors contributing to the noise often contain physically relevant information such as quantitative information about conformational degrees of freedom \cite{SPAfilter}.
  The abundance of data available to researchers and recent advances in computational and statistical methods are allowing researchers
to entertain new methods of summarizing information relevant to
 modeling systems at the nanoscale \cite{SPAfilter,SPAfric,gAIoan}.

 By applying surrogate models to the data coming from  biased MD simulations of DHFR,  it was  demonstrated that a collection of
  stochastic dynamical models can be used to better understand the factors contributing to the shape of the autocorrelation function associated
  with fluctuations coming from multiple time scales.  The surrogate models were estimated by appealing to maximum likelihood type
  methods \cite{aitECO,ozaki,chen09} and were tested using goodness-of-fit tests which utilized the transition density of the assumed surrogate
   and were appropriate for the data.  For example, the time series data  was not assumed to be stationary; the stationarity assumption is often
    suspect in simulation data.  The tests used \cite{hong} indicated that taking the position
dependence of the noise into account was required to provide a
statistically acceptable model in many regions of phase space
explored.  For short timescales, the individual surrogate models (taking position dependence noise into account)
were capable of predicting quantities outside of the fitting
criterion, e.g.  a parametric likelihood function was fit but the
models were able to predict short timescale autocorrelation
functions and these physically based models were able to fairly
accurately summarize/model relaxation kinetics that a simple exponential
relaxation  could not.
Other enzymes systems have exhibited this type of behavior
\cite{SPAJCTC} and it is likely that future single-molecule
experiments will yield data possessing this  feature.

 Perhaps more importantly, we demonstrated that a population of surrogate models was required to represent the complex dynamical system because an unobserved conformational degree freedom modulated the dynamical response and this ``random force" had to be accounted for in order to predict autocorrelations valid for longer temporal trajectories.  A method using parametric surrogate models calibrated over short timescales  while at the same time respecting the variability induced by unresolved coordinates evolving over longer  timescale was presented.  The DHFR system  was another instance where aggregating a collection of simpler dynamical  models gave rise to a more complex stochastic process \cite{granger80a,granger80b,cox91,SPAJCTC,klafter09}.     The basic idea is applicable to situations where a hidden slowly evolving degree of freedom modulates the dynamics and this coordinate evolves on an effective free energy surface possessing relatively low barriers \cite{SPAJCTC}.  Issues associated with extensions were  briefly discussed.

Even if a coarse system description, such as a single
autocorrelation function,   can be used to adequately  approximate
the physically relevant statistical properties of all
experimentally accessible observables,  the approach presented
still has appeal.
 One circumstance where this is particularly relevant is when  computer simulation trajectories are compared to frequently sampled experimental single-molecule time series \cite{SPAfilter}.  In experimental  time series, many conformational coordinates cannot typically be resolved \cite{SPAfilter,SPAJCTC},  so constructing a simulation that matches all relevant degrees of freedom is highly problematic.
 Quantitative  knowledge of how the variability induced by such hidden degrees of freedom  is reflected in the surrogate model parameters distribution may help in refining force fields to match kinetic properties at multiple timescales.  If the force fields are believed valid, then turning to the simulations for details of the structural dynamics can help us in understanding complex molecular machines \cite{schultenscireview}.  This type of extra detail may also assist (or lead to new) methods for computing transition rates \cite{chandler00}.
  Furthermore, as nanotechnology demands higher resolution at smaller length and timescales, one
   may want to avoid using  a single autocorrelation function constructed by aggregating many meso or microscopic states each possessing
   different dynamical features because doing so may unnecessarily wash out physically relevant information.
   The phenomenologically motivated simple bottom-up strategy  presented was one contribution in this direction.

\section{Acknowledgements}
The author thanks Karunesh Arora and Charles Brooks III for sharing the DHFR data.

\bibliographystyle{unsrt}
\bibliography{dhfr}

\section{Appendix }
\footnotesize
\subsection{Toy Model Parameters}

$\delta t=0.15, \delta s=\delta t/50, \mathcal{T}=200\times\delta t,  \tau_0=120,   (\alpha^0,\kappa^0,\eta^0)=(4,0.2,0.5), (\sigma^{\alpha},\sigma^{\kappa},\sigma^{\eta})=(6.5\times 10^{-2},6.5\times 10^{-3},1.9\times 10^{-2})$.  The last set of parameters were selected to give the evolving OU parameters a stationary distribution characterized by three independent normals each having mean  $(\alpha^0,\kappa^0,\eta^0)$ and standard deviation (1/2,1/20, 3/20).  The initial condition of each $y$ process was set to $\alpha^0$ and the OU parameters were all set to  $(\alpha^0,\kappa^0,\eta^0)$.  100 batches of 4 independent Brownian motion processes were used to evolve the system.

\subsection{Predicting Quantities with Surrogate Models}
 \label{sec:standardcomp}
%cite autocorrelation of OU.  nonlinear and
%regime switching (moments computable of polynomial...?cite{shoji_obscureREF}?})
The OU process is attractive for a variety of reasons.  Its  conditional and stationary density are both known analytically and it  can be readily estimated from discrete data.  For parameters possessing a stationary distribution, these can all be written explicitly in terms of  Normal densities.
Another appealing feature is that the AC function, denote this function by $AC(t)$, \cite{socci96} associated with a stationary process can readily be computed after parameter estimates are in hand, namely $AC(t)= \exp{-B t}$; recall the drift of the OU process is given by $B(A-z)$.

 Unfortunately, these types of statistical summaries   are more difficult to obtain with other SDEs.   Position dependence of the diffusion function and nonlinear models severely complicate obtaining  analytic expressions for the autocorrelation function.  Note that once a single SDE models is estimated, a new large collection of sample paths can be simulated and  quantities like the autocorrelation function associated with a given SDE model and $\theta$ can be empirically determined (the computational cost of simulating a scalar SDE is typically marginal in relation to a MD simulation).  This can be repeated for each surrogate SDE estimated from each MD path.

 However,  a stationary density,
under mild regularity conditions,  of a scalar SDE can often be expressed in closed-form  using only information contained in the estimated SDE coefficient functions via the relation \cite{kutoyants,risken} %pg 98 risken
\begin{equation}
p^{SD}(z;\Gamma)=\mathcal{Z}/\sigma(z)^2 \exp \Big( \int^z_{z_{\mathrm{REF}}}   \mu(z')/\sigma(z')^2  dz' \Big),
\label{eq:invardens}
\end{equation}
where in the above the SDE functions' dependence on $\theta$ and $\Gamma$ has been suppressed to streamline the notation.   $\mathcal{Z}$ represents a constant to ensure that the density integrates to unity and $z_{\mathrm{REF}}$ represents an arbitrary fixed reference point.  When evaluating $p^{SD}(\cdot)$, one  can encounter technical difficulties  if the diffusion coefficient is allowed  to take a zero or negative value (this is  relevant to the PSOD model). Some heuristic computational approaches to dealing with this are discussed in Refs. \cite{molsim,SPAJCTC}.

Sometimes a  thermodynamic motivation exists for expressing the stationary density of the high-dimensional molecular system in terms of some potential, denoted here by $V(z)$, that does not explicitly depend on the diffusion function \cite{risken,dima}. In time-homogeneous scalar overdamped Brownian dynamics, where the forces of interest acting on $z$ are believed related to the gradient of $V(z)$, a ``noise-induced drift" term \cite{risken} can be added to the drift function and this addition  cancels out the contribution coming from the  $1/\sigma(z)^2$ term outside the exponential.  The stationary density of the modified SDE can then be expressed as being  proportional to   $\exp{\big(-V(z)\big)}$ in such a situation.  This type of modification has a thermodynamic appeal when $z$ is the only important variable of the system and the fast-scale noise has been appropriately dealt with \cite{razstuart04}.  The utility of such an approach in describing the \emph{pathwise} kinetics of trajectories is another issue and single-molecule studies are one area where the distinction may be important (one may not care as much about the stationary \emph{ensemble}  distribution) .

 However when there are slowly evolving lurking variables like $\Gamma$ modulating the dynamics (as is the case in many biomolecular systems) using simple expression like Eq. \ref{eq:invardens} to approximate the stationary density of the high-dimensional system (with or without ``noise-induced drift" corrections) is highly problematic.
 Note that  the $\Gamma$ variable has been retained in the left hand side of Eq. \ref{eq:invardens};  the stationary density estimate is only meant to be valid for a fixed estimated SDE surrogate corresponding to one value $\theta$.  In this paper and others, it is assumed  that for a short time interval both $\theta$ and $\Gamma$ are effectively frozen.  Given a model and short time data, this can be  tested using goodness-of-fit tests.  However over longer timescales,  $\Gamma$ evolves and modulates the dynamics so the estimated  $\theta$ evolves in time (this is why the situation can be though of as  a type of dynamic disorder \cite{xie_dynamicdis_98}).  For this long time evolution, it is assumed that the form of a stochastic process depending only on  $z$ is completely unknown to the researcher.
 Furthermore it was assumed that another order parameter (i.e. system observable) is unavailable or is unknown \cite{SPAJCTC,SPAfilter,karplus08}.  Hence to approximate the stationary distribution of the high-dimensional molecular system one would require a collection of $p^{SD}$'s (each with different $\Gamma$'s) to approximate this quantity.  This procedure is presented in Ref.  \cite{SPAJCTC}.
 %A result along this line is presented for the DHFR system and its relevance to computational efficiency is %discussed in the Results.

\end{document}